\documentclass[12pt,a4paper]{article}
\begin{document}

\begin{center}
\begin{large}
NeXus Software Status\\
\end{large}
Mark K\"onnecke, Uwe Filges\\
Laboratory for Neutron Scattering\\
Paul Scherrer Institute\\
CH--5232 Villigen--PSI \\
Switzerland
\end{center}

\leftline{\large{Abstract}}
NeXus is a joint effort of both the synchrotron and neutron scattering
community to develop a common data exchange format based on HDF. In
order to simplify access to NeXus files a NeXus-API is provided. This
NeXus-API has been restructured and expanded to cover both HDF
versions 4 and 5. Only small changes to the API were necessary in
order to accomplish this. A foundation was laid to extend the
NeXus-API to further file formats. A new NeXus-API for IDL based on
the IDL C interface has been implemented. Thus both HDF-4 and HDF-5
NeXus files can be processed from IDL. The time-of-flight data
analysis program IDA has been adapted to support NeXus. The neutron
ray tracing package McStas from Risoe has been updated to write NeXus
files through the NXdict-API.

\section{Introduction}
NeXus$^{1}$ aspires to become a common data format for both the
synchrotron and neutron scattering community. The aim of the NeXus
proposal is to provide an efficient and platform independent self
describing data exchange format. The NeXus proposal has five levels:
\begin{itemize}
\item A physical file format.
\item A Application Programmer Interface (API) for accessing data
files.
\item A file structure.
\item Rules for storing single data items.
\item A dictionary of parameter names. 
\end{itemize}

As physical file format the hierachical data format (HDF)$^{2}$ from the
National Center for Supercomputing Applications (NCSA) was
choosen. This is a binary, platform independent, self describing data
format. HDF is well supported by major data analysis packages. HDF
files are accessed through a library of access functions. This library
is public domain software. Libraries are available for the ANSI--C,
Fortran77 and Java programming languages on all important computing
platforms. The HDF--4 library supports a lot of different content types
such as scientific datasets, groups, images, tables etc. Of these
NeXus only uses the scientfic dataset and the group structures. HDF--4
groups allow to order information hierarchically in a HDF file, much
like directories in a filesystem. As the
HDF library is also fairly complex a NeXus-API was defined which
facilitates  access to NeXus HDF files. 

The file structure part of the NeXus definition provides application
programs with the information where to find certain data elements in
the file. NeXus files are structured into several groups, much like
directories in a file system. The NeXus file structure provides for
multiple datasets in a single file and easy retrieval of plottable
information. NeXus encourages users to include all necessary
information about an experiment in one file and not to distribute such
information across multiple file. Therefore the NeXus file structure
provides for a complete description of an instrument used for an
experiment.

The rules for storing individual data items in a NeXus file provide
the infrastruture for locating the axises of multi--dimensional
datasets and require the user to store the units of measurement with
each data item.

The NeXus dictionary, the least developed area of the NeXus standard,
provides names for data items in a NeXus files and files structures
for known instrument types. More information about NeXus can be found
at the NeXus WWW-sites: 
\begin{center}
http://www.neutron.anl.gov/nexus and http://lns00.psi.ch/NeXus.
\end{center}

\section{A New NeXus-API}
The original NeXus-API as described above had been developed to
support the then prevalent HDF version 4.1 (HDF--4). Over time the HDF--4
library had become overly complex and also imposed certain limitations
on users. Therefore the NCSA decided to redesign HDF. This brought a
new version of HDF, HDF version 5, into existence which has a
different access library and a different, incompatible  file format.
HDF version 5
(HDF--5) maintained all the advantages of the HDF version 4 format. But
a couple of important limitations of HDF--4 were levied:
\begin{itemize}
\item HDF--4 limits the number of objects in a HDF file to 20000. This
sounds much, but isn't because most HDF-4 content types consist of
multiple objects.
\item File size was restricted to 2GB.
\item HDF--4 is not thread safe.
\end{itemize}
Some of these limitations of HDF--4 were already hit by NeXus
users. Though NCSA commited itself to maintaining HDF--4 for
some time to come, the need was felt to move to HDF--5 as the physical
file format for NeXus. 

Therefore the NeXus design team decided to program a new API with
following design goals:
\begin{itemize}
\item Support for both HDF--4 and HDF--5.
\item API compatibility to the old version.  
\end{itemize}
In order to achieve  this goal a framework was
developed for adding further file formats, for example XML, to the NeXus-API.

After some trouble with the HDF--5 libraries the new NeXus--API
version 2.0 was released. This version can be built to support either
HDF--4, HDF--5 or both. The goal of API compatibility was achieved
with three exceptions:
\begin{itemize}
\item Due to  technical reasons, the mechanism for the creation of
compressed datasets  had to change.
\item When creating a new file a user has to select if a HDF--4 or
HDF--5 is to be created.
\item In HDF--4 groups are ordered in the sequence of their creation
in the data file. In HDF--5 an alphabetical order is imposed. Thus the
order of groups in HDF--4 and HDF--5 NeXus files is not identical. The
HDF--5 team at NCSA promised to do something about this.   
\end{itemize}
A little problem was posed by NeXus classes. NeXus uses the class
name of a group and its name for identification. In HDF--5 the concept
of a group class was abandoned. However, HDF--5 brought us arbitrary
attributes to groups. The problem was thus solved by storing the NeXus
class in a group attribute with the name NXclass.  
So, the news is that there is no news! The NeXus-API stayed the
same. The new API is stable and is already in production use at the
SINQ--instruments TRICS and AMOR at PSI. 

The new NeXus--API is available for the programming languages:
\begin{itemize}
\item ANSI--C and C++
\item Fortran 77
\item Fortran 90
\item Java
\end{itemize}
It has been tested on True64Unix 5.1, Linux, Mac-OS 10 and Windows operating
systems so far. 

It is now recommended to use the newer HDF--5  file
format wherever possible.  Conversions from HDF--4 to HDF--5 and back
can be performed with a set of utilities provided by NCSA. This
conversion does not produce valid NeXus files though. The NCSA utility
for converting from HDF--4 to HDF--5, h4toh5, was made NeXus compliant
with a little patch. This utility also converts links properly. The
corresponding utility h5toh4, for conversion from HDF--5 to HDF--4
files,  can not be that easily patched to work with NeXus because
h5toh4 changes the file structure. There also exists a Java utility
for conversion from HDF--4 NeXus files to HDF--5 NeXus files and
back. This utility cannot be used in production use, though, because
it duplicates linked datasets.

\section{A New NeXus IDL API}
There has always been a NeXus--API for RSI's Interactive Data Language
(IDL) data treatment program. This NeXus--API for IDL (NIDL) was based
on the HDF--4 access functions provided by IDL. As HDF--5 support is not yet
available for IDL, this approach was not feasible for the new
NeXus--API. Moreover, a reimplementation using IDL HDF--5 access
functions would have meant a complete reimplementation and duplicated
code. Therefore it was choosen to integrate the new NeXus--API into
IDL through IDL's native function ANSI--C interface. This would also
save work in the case of further extensions to the NeXus file format
list.  This approach, a IDL NeXus API through IDL's native functions,
has been implemented. It currently is known to work well under
True64Unix version 5.1. As IDL standardizes many aspects of the native
function interface ports to other unix like operating system should
not be a big deal, a port to Microsoft Windows type operating systems may
require only a little more work.    

\section{New NeXus Aware Software}
\subsection{Browsers and Utilities}
Browsers allow to view the content of a NeXus file and sometimes even
edit it. The simplest of such applications is the command line browser
NXbrowse which is distributed with the NeXus package. It can browse both
NeXus HDF--4 and NeXus HDF--5 files.

Very new is the HDFView$^{3}$ program from NCSA. This is a merged version
of the older jhv and h5view applications from NCSA. HDFview
can display the hierarchy of a NeXus file and give some basic
renditions of data. HDFview also allows for editing of HDF
files. HDFview works with both HDF--4 and HDF--5. HDFview is written
in Java and thus available for a variety of platforms.

There is also a new NeXus Explorer program written in Visual Basic for
Windows by Albert Chan at ISIS. Besides browsing, this application can
also edit the NeXus file. With the help of IDL the application is
able to do 2D plots of data. The NeXus Explorer is only available for
the Windows platform from
\begin{center}
 http://www.isis.rl.ac.uk/geniebinaries/NexusExplorerBits.zip
\end{center} 

A couple of new NeXus utilities have been developed and been included
into the NeXus--API package:
\begin{description}
\item[NXtree] Manuel Lujan from APNS provided NXtree which displays
the  structure of a NeXus file as a tree. Output
can be in text, html or latex format.
\item[NXtoXML] is an experimental utility which converts a NeXus
binary file into a XML file. The xml format has still to be finalised
through.
\item[NXtoDTD] is another experimental program which documents the
structure of a NeXus program as a XML data Type definition (DTD). This
tool shall help in the process of defining the NeXus dictionary and
instrument type specifications. Both XML utilities were contributed by
Ray Osborn, APNS.  
\end{description} 

\subsection{Data Analysis Programs}

\subsubsection{General Data Analysis Programs}
Besides the known NeXus supporting data analysis tools IDL, opengenie
and Lamp there is now a new system named openDave brought to us by the
FRM--2 reactor group in munich. OpenDave has a modulare architecture
which allows to process data from a selection of sources through
various filters and output them to various types of sinks, including
graphics. OpenDave is written in C++ with the QT toolkit and is thus
restricted to operating systems supported by QT. Unfortunately, QT
requires a developers license on a lot of platforms 

\subsubsection{TOF Data Analysis Programs}
Besides the programs inx, nathan and Isaw, the IDA program from Andreas
Meyer, TU munich was adapted to support NeXus files as produced by the
FOCUS instrument at PSI. 

\subsubsection{Filters to Other Packages}
In this section filters to and from NeXus to other data formats are
discussed. For the small angle scattering community two such tools
exist:
\begin{description}
\item[psitohmi] converts PSI--SANS NeXus files to a format suitable
for the SANS data processing suite from the Hahn Meitner Institute in
berlin. This utility was provided by Joachim Kohlbrecher, PSI.
\item[nx2ill] Ron Gosh wrote nx2ill in order to convert SANS NeXus files
to the  format understood by the ILL SANS data processing tools. 
\end{description}

There exists also a small utility which combines powder diffraction
NeXus data files into a large powder diagram which is then stored in a
format suitable for the Rietveld program fullprof.

\subsubsection{Single Crystal Diffractometer Data Analysis}
At PSI a new program for the integration of single crystal diffraction
data collected with a PSD with the name {\bf anatric} was
written. This  program is
optimised for the SINQ single crystal diffractometer TRICS. TRICS is a
four circle diffractometer with a conventional eulerian cradle and
three position sensitive detectors positioned on a movable detector
arm at 0, 45 and 90 degrees offset. Typical measurements involve omega scans
across a given omega range with all other angles fixed. TRICS saves
its data in HDF--5 NeXus files. 

Anatric is able to perform two
operations:
\begin{description}
\item[Reflection location] anatric locates reflections in the data 
 without prior knowledge
about the crystal through a local maximum search. Reflection
positions are determined through a center of gravity calculation. The
output of this step is a list of reflections to be used for indexing
or UB matrix refinement. 
\item[Reflection integration] anatric can integrate reflection
intensities for further processing with a crystal structure data
determination package.  
\end{description}

Anatric has been designed to cope with two TRICS specific
problems. The first is that measurements are performed at low and at
high two theta at the same time. This together with the resolution of
the instrument  has the consequence, that
reflection positions shift by up to 2.5 pixels between omega frames
at the  high two theta detector. Moreover at TRICS reflections show up
as rather large features  on the detector. 
These two factors combined would make rectangular reflection
boxes for integration excessively large and thus impractical. Anatric
now takes this shift into account and extracts data to integrate along
an arbitrary axis through the reflection. This axis is determined
experimentally from strong reflections in the reflection location
step.

The second problem is the large size of the reflections at TRICS 
in relation to the size of the position sensitive detectors
(~20x20cm). A reflection box for integration must be large enough to
enclose strong reflections completely. Using such a large reflection
box for all reflections however would kill off many perfectly well
measured smaller reflections at the border of the detectors due to the
necessary border tests. In order to cope with this, anatric determines
for each reflection the size of the necessary integration box
individually. If the reflection is to weak for this to work, a minimum
integration box is used.  

The actual integration of intensities is then performed with a
variation of the dynamic mask procedure as described by Sj\"olin and
Wlodawer$^{5}$. Anatric has been written in C++. It is more or less
finished, however the program still needs to be verified against a full
structure refinement. 

\subsection{Monte Carlo Simulation Packages}
Monte Carlo simulations of instruments become an increasingly valuable
tool for instrument design, instrument optimisation and measurement
planning. The NeXus file format can offer full documentation of the
simulation run and efficient transfer of large simulated datasets to
such packages. Moreover it could also become possible to simulate data
files for testing of data analysis programs. Currently only one Monte
Carlo simulation program, McStas$^{6}$, has been ported to use NeXus for
data file storage by Emmanuel Farhi at ILL. As of now, only a XML
format is supported but support for binary NeXus files will follow
soon. Such binary NeXus files will be written through the NeXus
dictionary API. This allows the user to customize the NeXus file structure
to his demands. NeXus support for the simulation package Vitess$^{7}$ is in
early stages. 

\section{Future Developments of NeXus}
Most effort in developing the NeXus standard should now go into the
refinement of the dictionary and the definition of NeXus file
structures and content for a couple of different instrument types. It
would also be good to set up a kind of scheme for standardizing such
definitions. 

The other thing which needs to be done is to make the NeXus
installation and linking easiser. Feedback from users show that many
have difficulties when juggling with all the different libraries, 
NeXus, HDF--4 and HDF--5, when compiling and linking
programs. Possible solutions could be autoconf scripts for platforms
which support this and perhaps the automatic generation of a shell
script for linking against NeXus. 

One main objection against NeXus is: it is not ASCII, I cannot edit my
data! A possible answer to this objection would be a XML NeXus
format. XML means eXtended Markup Language and is a scheme for
defining a markup language in ASCII. The best known example of a
markup language is html which is driving the WWW. A XML NeXus format
is in the process of being defined. It is suggested to extend the
NeXus--API to support XML as well.

It would also be helpful to provide NeXus support for some more
general data analysis packages, especially free ones such as octave
and scilab. 

\section{Conclusion}
With the inclusion of the new HDF--5 file format into the NeXus--API,
this API is well braced for the future. More and more NeXus aware data analysis
software is being written and the software already available 
covers a wide area of applications. 

\section{Acknowledgements}
The port of the NeXus--API to HDF--5 was supported through the
european SCANS-network program. The SCANS RTD network was supported
under the Access to Research Infrastructure activity, Improving Human
Potential Program of the European Commission under contract number
HPRI--CT--1999-50013 and the Bundesamt f\"ur Bildung un Wissenschaft,
Berne, Switzerland under contract number BBW Nr.: 00.0208. This support is
gratefully acknowledged.  

\section{References}
\begin{enumerate}
\item P. Klosowski, M. Koennecke, J. Z. Tischler and R. Osborn: NeXus:
A common data format for the exchange of neutron and synchrotron
data. Physica B {\bf 241--243}, 151--153, (1998)
\item HDF: http://hdf.ncsa.uiuc.edu
\item http://hdf.ncsa.uiuc.edu/hdf-java-html/hdfview/index.html
\item openDave, 
\item L. Sj\"olin, A. Wlodawer, Improved Technique for Peak
Integration for Crystallographic Data Collected with
Position--Sensitive Detectors: A Dynamic Mask Procedure, Acta
Cryst. A37, 594--604, (1981)
\item K. Lefmann, K. Nielsen, McStas , Neutron News {\bf 10/3}, 20,
(1999)
\item D. Wechsler, G. Zsigmond, F. Streffer, F. Mezei: VITESS: Virtual
Instrumentation Tools for Pulsed and Continuour Sources, Neutron News
{\bf 11/4}, 25--28 (2000)
\end{enumerate}

\end{document}